\documentclass[11pt,a4paper,onecolumn]{journal}
\usepackage{amsmath,amssymb,amsthm}
\usepackage{graphicx, epsfig, upgreek}
\usepackage{hyperref}
\usepackage[left=1.in,right=1.in,top=1in,bottom=1in]{geometry}
\usepackage{authblk}
\usepackage{todonotes, color} 

\begin{document}
\title{Optimum size of nanorods for heating application}

\author[1]{Gowrishankar Seshadri}
\author[1]{Rochish Thaokar}
\author[1]{Anurag Mehra}
\affil{Department of Chemical Engineering, Indian Institute of Technology Bombay, India}
\date{\today}
\maketitle

\begin{abstract}
Magnetic nanoparticles (MNP's) have become increasingly important in heating applications such as hyperthermia treatment of cancer due to their ability to release heat when a remote external alternating magnetic field is applied. It has been shown that the heating capability of such particles varies significantly with the size of particles used. In this paper, we theoretically evaluate the heating capability of rod-shaped MNP's and identify conditions under which these particles display highest efficiency. For optimally sized monodisperse particles, the power generated by rod-shaped particles is found to be equal to that generated by spherical particles. However, for particles which have a dispersion in size, rod-shaped particles are found to be more effective in heating as a result of the greater spread in the power density distribution curve. Additionally, for rod-shaped particles, a dispersion in the radius of the particle contributes more to the reduction in loss power when compared to a dispersion in the length. We further identify the optimum size, i.e the radius and length of nanorods, given a bi-variate  log-normal distribution of particle size in two dimensions.
\end{abstract}


\renewcommand {\baselinestretch} {1.1} \normalsize
\fontsize{11pt}{15pt}\selectfont

\section{Introduction}
\label{sec:intro}
A fascination for the study of MNP's arose due to the potential benefits of their small size coupled with their intrinsic magnetic properties. The synthesis of a cobalt cluster of nanoparticles was first reported in 1995  \cite{ref8}. Following this, several groups synthesized both isotropic \cite{ref1} and anisotropic nanoparticles \cite{ref2,ref3} through a variety of methods. The magnetic properties of these particles have also been reported by a range of studies on MNP's and clusters \cite{ref6}. \\

MNP's have been used in heating applications due to hysteresis and relaxational losses which lead to heat dissipation by these particles when present in suspensions \cite{ref7}. It has been shown conclusively that under an alternating magnetic field MNP's possess the ability to produce intense heat around a small, localized region \cite{ref24}. The localized heat generation produced by MNP's has gained considerable importance due to its utility in hyperthermia treatment of cancer\cite{ref16, ref15, ref14, ref13}. The temperature profile produced by such particles when present in tumours has been studied extensively through numerical and analytical techniques \cite{ref12,ref20, ref21}.  Various studies have examined the role of particle size, material of synthesis, frequency and amplitude of the applied magnetic field on the effective loss power produced by the particles. \cite{ref18, ref19, ref20,ref9}   \\

It has been shown through analytical and experimental work that the size of MNP's critically affects the power generated by them \cite{ref24, ref25, ref9}. For spherical MNP's, an optimum size has been found to exist for which the heat generation per unit volume (loss power density) of the nanoparticle is maximum. This occurs due to the parallel nature of the Neel and Brownian relaxation mechanisms of heat generation for nanosized superparamagnetic nanoparticles, where the Neel relaxation time increases with decrease in particle size and the Brownian relaxation time decreases with decrease in particle size \cite{ref11, ref4, ref27}. The use of anisotropic particles for hyperthermia application has been limited by the suggestion that a high magnetic field is necessary to completely utilize the magnetic hysterisis loop of such particles \cite{ref4}. However, such hysterisis heat loss is expected only for larger sized nanorods, and relaxational losses are expected for nanorods in a smaller size range. In this paper, we focus our attention on rod-shaped particles where the loss power would be determined primarily through relaxational losses. This analysis is based on experimental work on the heating effects of anisotropic nanoparticles which highlights a potential use of such particles in hyperthermia treatment \cite{ref25, ref10, ref5}. Further, we extend the theoretical understanding of heat generation to anisotropic MNP's and show that the maximum power density produced by optimally sized monodisperse rod-shaped and spherical-shaped nanoparticles is equal. Additionally, we find that when these particles are present with a similar distribution in size, the power generated by nanorods is dramatically enhanced when compared to that by spherical particles. The larger heat generation by rod-shaped particles compared to spherical particles when both are present as a distribution in a solution presents a potential advantage for the use of rod-shaped MNP's in hyperthermia treatment by reducing the dosage of material necessary. We further propose that this enhancement of heat generation in nanorods can be used as a measure to predict the size distribution of magnetic particles present in a solution, in addition to providing a crude estimate of any shape defects of particles present in a solution.  \\

\section{General Formulation}
\label{genform}
\subsection{Heat Generation through Relaxational Losses}
In our study, we focus our attention on the heat generated by nanoparticles through relaxational losses and ignore the losses due to hysteresis. For small sized nanoparticles, heat loss is governed mainly by two relaxation mechanisms : the Neel relaxation mechanism and the Brownian relaxation mechanism. Neel relaxation occurs  due to flipping of the orientation of the domain magnetic moment of nanoparticles with respect to the external field in a finite time, known as the Neel relaxation time ($\tau_n$). In the Brownian relaxation mechanism, heat is generated by the rotation of the entire particle in the magnetic field, through Brownian rotation of particles. The mean time for a finite rotation of the particle is known as the Brownian relaxation time ($\tau_b$).  Heat loss through Neel relaxation cycles are predominant for smaller sized nanoparticles while Brownian relaxational cycles are favoured by larger sized nanoparticles. For MNP's, the heat generated through these two mechanisms occurs in a parallel manner, since any isolated nanoparticle can follow either of two modes of heat loss in a finite duration of time. Therefore, the overall time constant for heat generation is given by \cite{ref4}: 
\begin{equation}
\tau=\tau_b^{-1}+\tau_n^{-1}
\end{equation}
\begin{equation}
\tau_b=\frac{1}{2D_r} \, \tau_n=\frac{\sqrt{\pi}\tau_0\exp{\Gamma}}{2\sqrt{\Gamma}}
\end{equation}
Here $\tau_0$ is the characteristic relaxation time and $\Gamma=KV_m/k_bT$ is a measure of the anisotropic energy of these particles when compared to their thermal energy. $k_b$ is the fundamental Boltzmann constant, $T$ is the temperature of the solution and $K$ is the anisotropy constant of the particles, the effect of which is studied later in the section. \\

The Brownian relaxation time ($\tau_b$) is directly proportional to the viscosity ($\eta$) of the medium and thus for highly viscous media, nanoparticles take a longer time to rotate physically. For colloidal particles the diffusivity ($D_r$) is given by $D_r=f_r/k_bT$, through Einstein's formulation, where $f_r$ is the friction factor of the colloidal system. \\

In our case, the corresponding friction factor of interest is that due to the Brownian rotation of the colloidal nanoparticle system. Rod-shaped particles have been modelled as prolate ellipsoids, and the friction factor corresponding to axial rotation of particles is only considered due to lower moment of inertia associated with such a rotation compared to that around the perpendicular axis. We define a dimensionless friction factor as $F_r=\frac{f_r}{8\pi\eta R_e^3}$. For spherical particles, $F_r=1$ and $R_e$ is the radius of the sphere. For rod-shaped nanoparticles \cite{ref28}, 
\begin{align}
F_r&=\frac{4(1-q^2)}{3(2-2q^{4/3}/F_t)}\\
F_t&=\frac{\sqrt{1-q^2}}{q^{2/3}\ln{\left(\frac{1+\sqrt{1-q^2}}{q}\right)}}\\
R_e&=(lr^2)^{1/3}
\end{align}
Here,$q=1/a$ where $a$ is the aspect ratio of the rod-shaped particle, which is defined as the ratio of the diameter of the ellipsoid ($2r$) to the length of the ellipsoid ($l$). Similar expressions described in \cite{ref28}can be used to determine the translational friction factor as well as the rotation friction factor about the perpendicular axis.  \\

Based on these expressions developed, the overall heat generation due to magnetic hysteresis is given by \cite{ref4} :
\begin{equation}
\label{equation:power}
P=\pi \mu_0 \chi_0H_0^2f\frac{2\pi f\tau}{1+(2\pi f\tau)}
\end{equation}

Here $P$ is the power generated per unit volume of the nanoparticle, $\mu_0$ is the magnetic permeability of free space, $\chi_0$ is the magnetic susceptibility of the material, $H_0$ is the magnitude of the applied field strength, $f$ is the frequency of magnetic oscillations and $\tau$ is the mean relaxation time for the process. From Equation \ref{equation:power}, it is clear that when $f\tau \gg 1$, the power generated by the particles varies linearly with the external frequency of oscillations. When $f\tau \ll 1$, the power generated by the particles tends to $0$. It is also evident from the expressions, that the maximum power generated by the nanoparticles occures when $f=\frac{1}{\tau}$. The magnetic susceptibility of a nanoparticle $(\chi_0)$ is given by : 
\begin{equation}
\chi_0=\frac{\mu_0\phi M_d^2V_m}{3kT}\frac{3}{\zeta}\left(\coth{\zeta}-\frac{1}{\zeta}\right)\\
\end{equation}
Here $\phi$ is the volume fraction of the particles in the solution, $M_d$ is the domain magnetization of the particles. $\zeta=\mu_0M_dHV_m/k_bT$ where $H$ is the magnetic field which has been externally applied on the system. From these expressions, it is evident that the heat generated by MNP's present in a solution is volumetric in nature, and shape affects both the anisotropic constant $(K)$ for the system and the brownian relaxation time, albeit to a lesser degree. \\

Expressions derived above, for calculating the heat loss by spherical nanoparticles have been well studied \cite{ref6}. For rod-shaped particles, the effect of increased anisotropy of the particles is modelled by modifying the anisotropy constant $(K)$ of the particles present in the solution. For spherical particles, the anisotropic constant of the system is solely determined by the effect of various crystal plane orientations on the magnetic polarisability in each planar direction, which is known as the magneto-crystalline anisotropy constant, denoted by  $K_{mag}$. For rod-shaped particles, in addition to the magneto-crystalline anisotropy constant, a shape anisotropy also exists due to the non symmetric nature of rod-shaped particles, denoted by $K_{shape}$. The effect of both these anisotropy constants is taken in a cumulative manner to determine the overall anisotropy constant for rod-shaped particles, denoted by $K_{overall}$. Thus for rod-shaped particles, $K_{overall}=K_{mag}+K_{shape}$. The experimental study of heat generation by spherical particles can allow us to determine the magneto-crystalline anisotropy constant of the material. This contribution is assumed to be identical for both spherical and rod-shaped particles of the same material. Due to the symmetrical nature of spherical particles, the contribution of shape to the anisotropy constant for such particles is $0$. Based on the theory developed in Cullity,et al. \cite{ref23}, the shape anisotropy for rod-shaped particles can be estimated by approximating them as prolate spheroids, and is given by : 
\begin{equation}
\label{eq:kshape}
K_{shape}=0.5\mu_0(N_a-N_c)M_s^2
\end{equation}

Here, $N_a$ and $N_c$ represent the demagnetization constants for the rod-shaped particles and are functions of the aspect ratio ($a$). Rod-shaped particles are again modelled as prolate spheroids, and the demagnetization along the two directions is given by \cite{ref23} :
\begin{equation}
\label{eq:kshape2}
\begin{split}
&N_c=\frac{1}{(a^2-1)}\left[\frac{a}{\sqrt{a^2-1}}\ln{(a+\sqrt{a^2-1})}-1\right]\\
&N_a=N_b=\frac{1-N_c}{2}
\end{split}
\end{equation}

Based on these expressions, the value of $K_{shape}$ for rod-shaped particles can be evaluated. The corresponding loss power for both spherical and rod-shaped nanoparticles is calculated and compared in section~\ref{results}.

\subsection{Distribution of Particle Sizes}
Synthesis of nanoparticles often results in a distribution of particle with varying sizes being present in the solution. A non-uniform distribution of particles can affect the power generated by the solution, since the loss power density function is strongly dependent on the size of the particles present in the solution. In our paper, we report the standard deviation in the particle radius as a fraction of the mean particle radius and similarly the standard deviation in the length of the particle as a fraction of the mean particle length. A log-normal distribution of particle size is assumed based on previous experimental and theoretical work \cite{ref4}. The probability density function representing the fraction of rod-shaped particles, with a given radius and aspect ratio, is given by :
\begin{equation}
y(r,a)=\left[\frac{1}{r\sigma_r \sqrt{2\pi}}\exp{\frac{-(\ln{r}-\mu_r)^2}{2\sigma_r^2}}\right]\left[\frac{1}{l\sigma_l \sqrt{2\pi}}\exp{\frac{-(\ln{l}-\mu_l)^2}{2\sigma_l^2}}\right]
\end{equation}
Here, $\mu_r$ is the mean radius of the particle, $\mu_l$ is the mean length of the particles, $\sigma_r$ is the standard deviation in the radius of the particle and $\sigma_l$ is the standard deviation in the length of the particles. For the case of spherical particles, the probability density function representing the fraction of particles with a given radius is given by : 
\begin{equation}
y(r)=\left[\frac{1}{r\sigma_r \sqrt{2\pi}}\exp{\frac{-(\ln{r}-\mu_r)^2}{2\sigma_r^2}}\right]
\end{equation}
For such a distribution of particles, the power generated per unit volume varies based on the size and shape of individual nanoparticles present in the system. In general, if the power generated per unit volume of a particle is represented by $P(n)$, where  $n$ represnts the number of particles in a solution with a particular dimension, the average power delivered to the solution by a distribution of particles of various sizes is given by : 
\begin{equation}
P_{avg}=\frac{\int{P(n)dn}}{\int{dn}}
\end{equation}
For the purposes of our discussion, we will use reported values of the fundamental properties of $Fe_3O_4$ given in Table~\ref{table1} throughout the following section, to illustrate the key findings of our analysis \cite{ref4} . The table also shows the assumed heating conditions used in the simulation. Additionally, the results presented are independent of the concentration of nanoparticles in the solution, and the loss power of the solution is known to scale linearly with the concentration of particles in the system. 
\begin{table}
\centering
\caption{Table shows the assumed magnetic property values and heating conditions for $Fe_3O_4$ nanoparticles \cite{ref4}}
\label{table1}
\vspace{4mm}
\begin{tabular}{|l|l|}
\hline
Particle Property & Value \\ \hline
Saturation Magnetization & 470 $kA/m$ \\ \hline
$H_0$ & 6.5 $ kA/m$ \\\hline
Frequency & 100 kHz \\\hline
Magneto-crystalline Anisotropy Constant & 10 $kJ/m^3$  \\\hline
\end{tabular}
\vspace{-2mm}
\end{table}

\section{Results and Discussion}
\label{results}
\begin{figure}[htbp!]
\includegraphics[width=\textwidth]{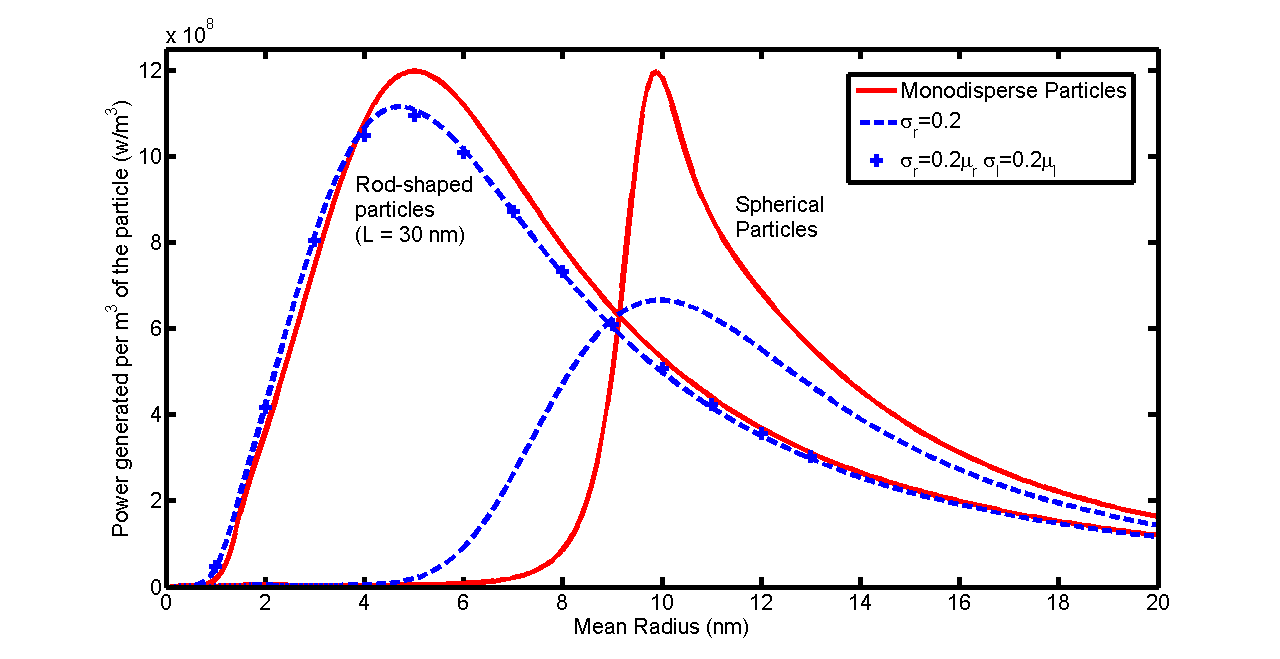}
\caption{Figure shows the variation of the loss power density in spherical and rod-shaped particles with a fixed length of 30 nm. The solid line shows the variation of the loss power density for monodisperse particles and the dashed lines show the variation of the loss power density for particles with a standard deviation in the radius equal to $20\%$ of the mean radius. The points $(+)$ show the loss power density of particles with a fixed standard deviation of $20\%$ of the mean  length in addition to a standard deviation equal to $20\%$ of the mean radius.  }
\label{fixedh}
\end{figure}

\begin{figure}[htbp!]
\includegraphics[width=\textwidth]{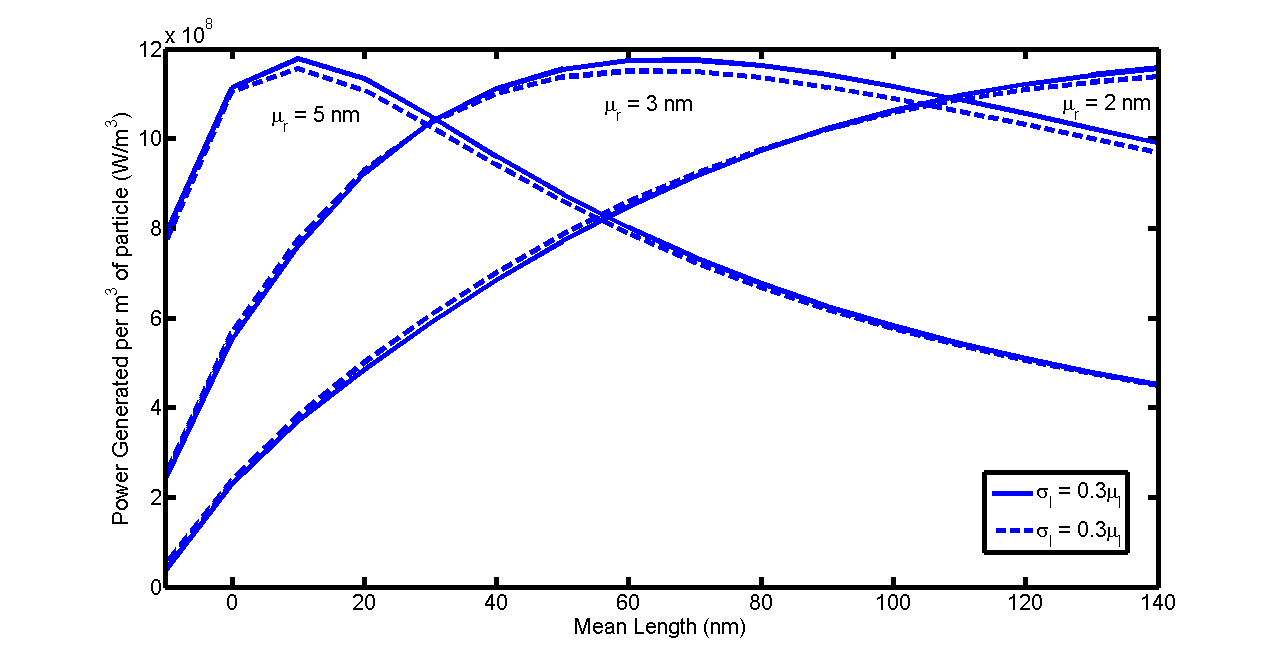}
\caption{Figure shows the variation of the loss power density in rod-shaped particles with differing radius as a function of its length with a dispersion only in the length of the particle. The solid line shows the variation for particles with a standard deviation equal to $20\%$ of the mean length and the dashed lines show the variation for particles with a standard deviation equal to $30\%$ of the mean length. }
\label{fixedr}
\end{figure}

Using expressions 1-6 developed in the previous section, the power generated by monodisperse particles can be calculated by assuming $K_{shape}=0$ for spherical particles and by calculating $K_{shape}$ for rod-shaped particles for a given aspect ratio using equations~\ref{eq:kshape} and \ref{eq:kshape2}. Further, the loss power of spherical particles present with a size distribution was compared to both that of rod-shaped particles with a distribution in the radius and a fixed particle length as well as rod-shaped particles with a distribution in both radius and length. This comparision is illustrated in   Figure~\ref{fixedh}. We find that while monodisperse rods and spheres have the same value of maximum power output, rod-shaped particles have the propensity to generate greater power when both the particles have the same distribution in the particle radius. In addition to this, we see that a shift in the position of the maxima also occurs for particles which are distributed. A higher maximum loss power density for rod-shaped particles with a distribution occurs because the peak for monodisperse rod-shaped particles is broader as compared to that of spherical particles. As a result for rod-shaped particles present with a size distribution, there is a greater contribution to the average loss power density from particles with dimensions close to that present at the maxima. Additionally, the effect of a variation in the length of rod-shaped particles is found to be negligible compared to the effect of a variation in particle radius, on the overall loss power density. The effect of a distribution in length is further illustrated in figure~\ref{fixedr}, where we study the effect of a distribution in the length of nanorods on the power generation term for particles with a fixed radius. We find that the effect of a variation in the length of nanorods on the power generation term only marginally decreases the loss power density of the particles. \\

\begin{figure}[htbp!]
\includegraphics[width=\textwidth]{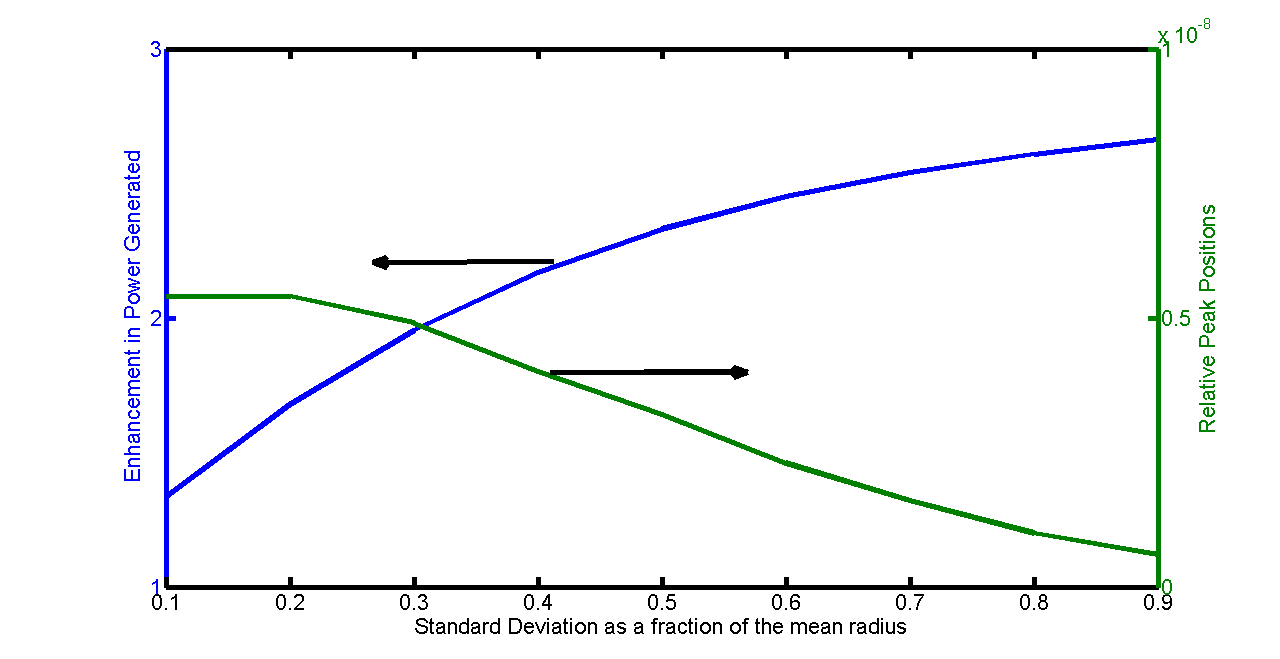}
\caption{Figure shows the enhancement in maximum power generated and the relative position of the maxima for rod-shaped particles (Fixed Length=30 nm) when compared to spherical particles when both particles have identical distributions in the radius with the standard deviation being represented as a percentage of the mean particle radius.}
\label{increasedgen}
\end{figure}

Since the effect of a variation in the length of nanorods on the overall power generated is small, we focus our study on the effect of a variation in the particle radius of nanorods  when compared to spherical particles.  The enhancement in the power generated is defined as the ratio of the maximum power generated by dispersed rod-shaped particles to that generated by dispersed spherical particles. In figure~\ref{increasedgen}, the enhancement in power generation, along with the relative peak position distance for rod-shaped and spherical-shaped nanoparticles is shown, for varying variance in the particle radius. A saturation in the enhancement of power generation occurs for large standard deviations, which is replicated for particles of varying fixed lengths. This saturation occurs due to the marginal contribution to the overall power by an increasing number of particles in the fringes, as the deviation in radius increases. A shift in the peak position (position at which the maximum power generation density is observed) of rod shaped particles occurs with respect to spherical particles occurs, and it has been found that the peaks shift closer for larger standard deviations in the particle size.  \\

\begin{figure}[htbp!]
\includegraphics[width=\textwidth]{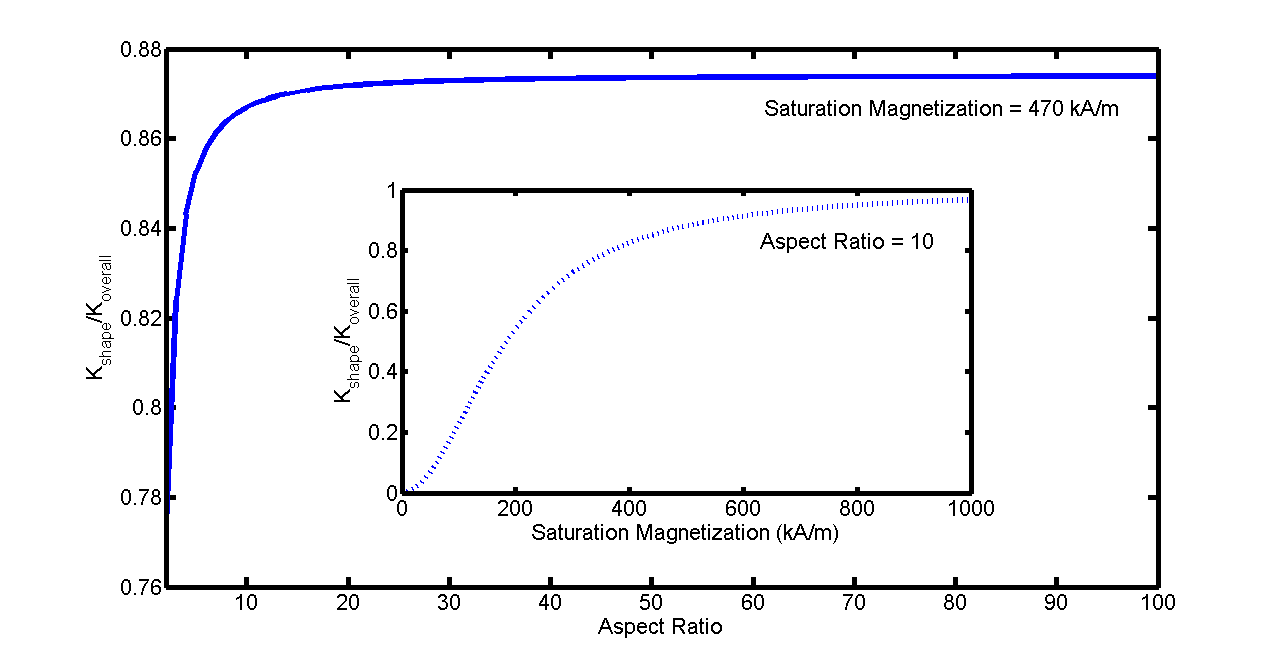}
\caption{Figure shows both the fractional contribution of $K_{shape}$ to the overall anisotropy constant as a function of the aspect ratio and for varying $M_s$ for particles with a fixed aspect ratio of 10 (inset figure). }
\label{shapecomp}
\end{figure}

We further investigate the fractional contribution of $K_{shape}$ to the overall anisotropic constant for varying aspect ratios in figure~\ref{shapecomp}. As described by equations~\ref{eq:kshape} and \ref{eq:kshape2}, $K_{shape}$ depends only on the aspect ratio of particles present in the system. We see that the shape contribution to the overall anisotropic constant saturates when particles with large aspect ratios are present. In figure~\ref{shapecomp}, we further illustrate the variation of the fractional contribution of $K_{shape}$ to the overall anisotropic constant for increasing values of the saturation magnetization of the suspension. For larger values of the saturation magnetization, a saturation in the fractional contribution of $K_{shape}$ to the overall anisotropy constant occurs. We note that an addition of a small degree of anisotropy to particles alters the anisotropy constant quite significantly, thus implying that even a small degree of anisotropy can help in reducing the impact of a distribution on the loss power density for spherical particles. 

\begin{figure}[htbp!]
\includegraphics[width=\textwidth]{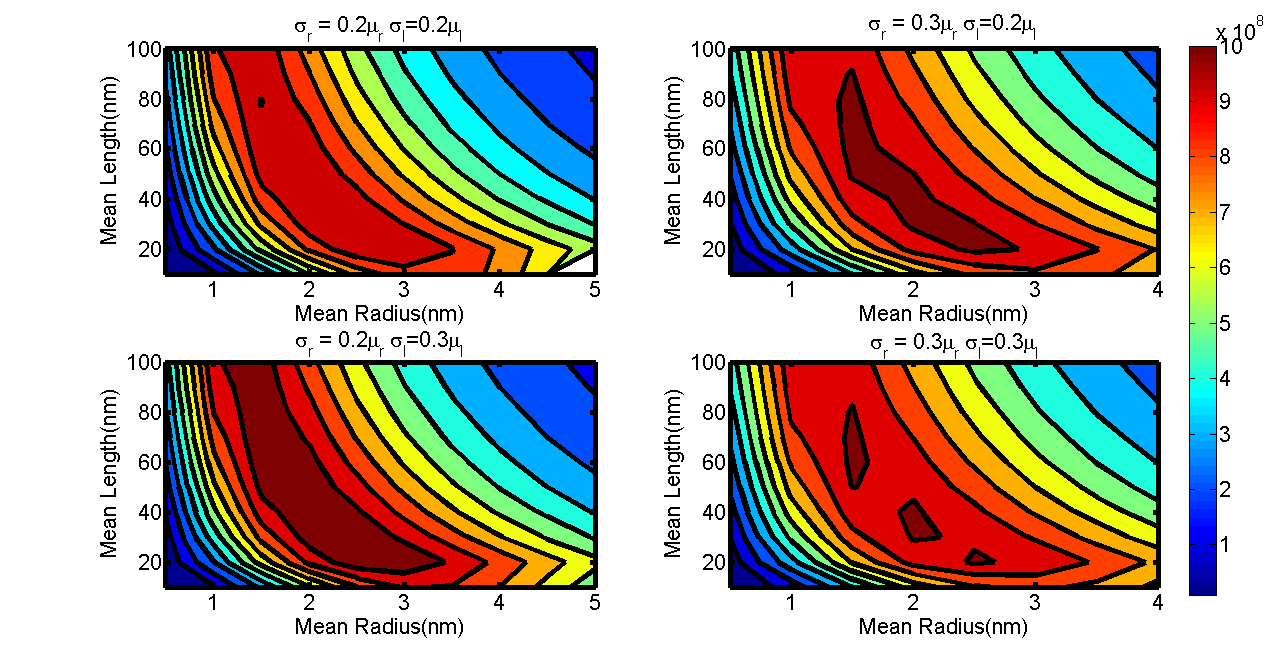}
\caption{Figure shows the power generation density for bi-variately distributed rod-shaped particles in both the length as well as particle radius where the standard deviation is represented as a fraction of the mean size. }
\label{contourplot}
\end{figure}

The optimum size map for rod-shaped particles which have a bi-variate distribution in the length as well as the radius is also plotted. This result portrays a realistic case for rod-shaped particles since it represents a distribution in both the dimensions of the rod.  Figure~\ref{contourplot} shows the power generated based on a bivariate distribution in the particle size for varying deviations in both dimensions. The figure gives us a representative map of the power generated for a range of particle radius and length. We find that varying the deviation in the length of the particle does not alter the contour plot, since its effect on the power generation term is small. Additionally, there exists a region of high power density, for a given set of distribution parameters for the system. This region of high density represents the ideal dimensions for rod-shaped nanoparticles which should be used for hyperthermia treatment. These plots can be constructed for MNP's of any material, given the magnetic properties of the material as shown in table~\ref{table1}, and the expected deviation in both dimensions which is based on deviation known to be produced through a given synthesis route. This method would enable us to design anisotropic nanoparticles with specific dimensions in order to obtain a high power density. It further allows us to characterize the distribution of nanoparticles given the mean radius and length of the synthesized particles. From figure~\ref{contourplot}, we see that particles with a smaller radius and a larger length are suitable for generation of a high power density. We further note that these maps could potentially be used to identify the distribution of nanoparticles in a solution with a known loss power density.

\section{Conclusion}
In this paper, the role of anisotropic rod-shaped particles in heat generation for application in magnetic hyperthermia has been studied. It has been shown that monodisperse rod-shaped particles have the same maximum power generation as monodisperse spherical particles despite the greater magnetic anisotropic energy of rod-shaped particles. However, the additional anisotropy present in rod-shaped particles increases the spread of the power distribution curve. As a result, for MNP's with a distribution, we find that the loss power density generated by rod-shaped particles is greater than that generated by spherical particles with the same deviation in size. Additionally, we have shown that for rod-shaped particles with a bi-variate distribution, an optimum region of maximum power density exists. Such knowledge can help us design optimally sized nanorods for heating applications given the distribution parameters which arise due to the chosen method of synthesis. It can also help us determine the distribution parameters of a system if the loss power density function is known.


\begin{thebibliography}{9}
\bibitem{ref1} Cordente, Nadege, et al. "Synthesis and magnetic properties of nickel nanorods." Nano letters 1.10 (2001): 565-568.
\bibitem{ref2} Gambardella, P., et al. "Giant magnetic anisotropy of single cobalt atoms and nanoparticles." Science 300.5622 (2003): 1130-1133.
\bibitem{ref3} Puntes, Victor F., Kannan M. Krishnan, and A. Paul Alivisatos. "Colloidal nanocrystal shape and size control: The case of cobalt." Science 291.5511 (2001): 2115-2117.
\bibitem{ref4} Hergt, Rudolf, et al. "Physical limits of hyperthermia using magnetite fine particles." Magnetics, IEEE Transactions on 34.5 (1998): 3745-3754.
\bibitem{ref5} Usov, N. A. "Low frequency hysteresis loops of superparamagnetic nanoparticles with uniaxial anisotropy." Journal of Applied Physics 107.12 (2010): 123909-123909.
\bibitem{ref6} Park, Sang-Jae, et al. "Synthesis and magnetic studies of uniform iron nanorods and nanospheres." JOURNAL-AMERICAN CHEMICAL SOCIETY 122.35 (2000): 8581-8582.
\bibitem{ref7} Hergt, R., et al. "Maghemite nanoparticles with very high AC-losses for application in RF-magnetic hyperthermia." Journal of Magnetism and Magnetic Materials 270.3 (2004): 345-357.
\bibitem{ref8} Gibson, Charles P., and Kathy J. Putzer. "Synthesis and characterization of anisometric cobalt nanoclusters." Science (New York, NY) 267.5202 (1995): 1338.
\bibitem{ref9} Chen, Qi, and Z. John Zhang. "Size-dependent superparamagnetic properties of MgFeO spinel ferrite nanocrystallites." Applied physics letters 73 (1998): 3156.
\bibitem{ref10} Lo, Chieh-Tsung, and Po-Yu Kuo. "Synthesis and Magnetic Properties of Iron Phosphide Nanorods." The Journal of Physical Chemistry C 114.11 (2010): 4808-4815.
\bibitem{ref11} Mehdaoui, Boubker, et al. "Optimal size of nanoparticles for magnetic hyperthermia: A combined theoretical and experimental study." Advanced Functional Materials (2011).
\bibitem{ref12} Andrä, W., et al. "Temperature distribution as function of time around a small spherical heat source of local magnetic hyperthermia." Journal of Magnetism and Magnetic Materials 194.1 (1999): 197-203.
\bibitem{ref13} Aqil, Abdelhafid, et al. "Magnetic nanoparticles coated by temperature responsive copolymers for hyperthermia." J. Mater. Chem. 18.28 (2008): 3352-3360.
\bibitem{ref14} Ito, Akira, et al. "Complete regression of mouse mammary carcinoma with a size greater than 15 mm by frequent repeated hyperthermia using magnetite nanoparticles." Journal of bioscience and bioengineering 96.4 (2003): 364-369.
\bibitem{ref15} Jordan, Andreas, et al. "The effect of thermotherapy using magnetic nanoparticles on rat malignant glioma." Journal of neuro-oncology 78.1 (2006): 7-14.
\bibitem{ref16} Berry, Catherine C., and Adam SG Curtis. "Functionalisation of magnetic nanoparticles for applications in biomedicine." Journal of physics D: Applied physics 36.13 (2003): R198.
\bibitem{ref17} Zhang, Chuanqian, Duane T. Johnson, and Christopher S. Brazel. "Numerical study on the multi-region bio-heat equation to model magnetic fluid hyperthermia (MFH) using low Curie temperature nanoparticles." NanoBioscience, IEEE Transactions on 7.4 (2008): 267-275.
\bibitem{ref18} Pavel, M., G. Gradinariu, and A. Stancu. "Study of the optimum dose of ferromagnetic nanoparticles suitable for cancer therapy using MFH." Magnetics, IEEE Transactions on 44.11 (2008): 3205-3208.
\bibitem{ref19} Pavel, M., and A. Stancu. "Study of the optimum injection sites for a multiple metastases region in cancer therapy by using MFH." Magnetics, IEEE Transactions on 45.10 (2009): 4825-4828.
\bibitem{ref20} Maenosono, Shinya, and Soichiro Saita. "Theoretical assessment of FePt nanoparticles as heating elements for magnetic hyperthermia." Magnetics, IEEE Transactions on 42.6 (2006): 1638-1642.
\bibitem{ref21} Liu, Kuo-Chi, and Han-Taw Chen. "Analysis for the dual-phase-lag bio-heat transfer during magnetic hyperthermia treatment." International Journal of Heat and Mass Transfer 52.5 (2009): 1185-1192.
\bibitem{ref22} Habib, A. H., et al. "Evaluation of iron-cobalt/ferrite core-shell nanoparticles for cancer thermotherapy." Journal of Applied Physics 103.7 (2008): 07A307-07A307.
\bibitem{ref23} Cullity, Bernard Dennis, and Chad D. Graham. Introduction to magnetic materials. Wiley-IEEE Press, 2011.
\bibitem{ref24} Park, Jongnam, et al. "Synthesis, characterization, and magnetic properties of uniform-sized MnO nanospheres and nanorods." The Journal of Physical Chemistry B 108.36 (2004): 13594-13598.
\bibitem{ref25} Gregg, Kristy A., et al. "Controlled synthesis of MnP nanorods: Effect of shape anisotropy on magnetization." Chemistry of materials 18.4 (2006): 879-886.
\bibitem{ref26} Rosensweig, R. E. "Heating magnetic fluid with alternating magnetic field." Journal of Magnetism and Magnetic Materials 252 (2002): 370-374.
\bibitem{ref27} Khandhar, Amit P., R. Matthew Ferguson, and Kannan M. Krishnan. "Monodispersed magnetite nanoparticles optimized for magnetic fluid hyperthermia: Implications in biological systems." Journal of applied physics 109.7 (2011): 07B310.
\bibitem{ref28} Bloomfield, Victor A. "Survey of biomolecular hydrodynamics." On-Line Biophysics Textbook: Separations and Hydrodynamics (2000).


\end{thebibliography}
\end{document}